\documentclass[preprint]{revtex4-1} 

\usepackage{fullpage}
\usepackage{amssymb}
\usepackage{amsmath}
\usepackage{setspace}
\usepackage{graphicx}
\usepackage{hyperref}
\usepackage{natbib}
\usepackage{afterpage}
\usepackage{float}

\begin{document}

\title{{\bf Observation of Deeply-Bound $^{85}$Rb$_{2}$ Vibrational Levels Using Feshbach Optimized Photoassociation}}
\author{S.~P. Krzyzewski, T.~G. Akin, J.~Dizikes, Micheal~A. ~Morrison, E.~R.~I. Abraham}
\affiliation{University of Oklahoma, Homer L. Dodge Department of Physics and Astronomy, Norman, OK 73019}
\date{\today}
\begin{abstract}

We demonstrate Feshbach optimized photoassociation (FOPA) into the $0_{g}^{-} (5$S$_{1/2}+5$P$_{1/2}$) state in $^{85}$Rb$_{2}$.
FOPA uses the enhancement of the amplitude of the initial atomic scattering wave function due to a Feshbach resonance to increase the molecular formation rate from photoassociation. 
We observe three vibrational levels, $v=127$, 140, and 150, with previously unmeasured binding energies of 256, 154, and 96 cm$^{-1}$.
We measure the frequency, central magnetic field position, and magnetic field width of each Feshbach resonance. 
Our findings experimentally confirm that this technique can measure vibrational levels lower than those accessible to traditional photoassociative spectroscopy.

\end{abstract}
\maketitle

\section{Introduction}

Molecular potential energy curves are the foundation of understanding atomic interactions.
Accurate knowledge of these potentials is essential to implement and analyze many experiments with cold and ultracold molecules \cite{T09}.
Cold molecules are important for a number of areas of interest, such as tests of fundamental symmetries, ultracold chemistry, quantum degenerate gases characterized by strong anisotropic dipole-dipole interactions, and quantum computation (\cite{KGP09} and references within).

The most accurate potential curves are those based on experimental determination of the vibrational energy level spectrum.
Traditional molecular spectroscopy provides much of this data, but is often limited to spectra near the minimum of broad molecular potentials.
Twenty years ago, the advent of laser cooling and trapping allowed  a new technique, photoassociative spectroscopy (PA), to access high-lying vibrational states of excited \cite{ LHP93, MCH93, JTL06} and ground \cite{AMS95} state molecular potentials.

While for a few systems these techniques overlap to provide a complete picture of the interatomic interaction \cite{MAH96}, for many others, a substantial gap exists.  
For example, for the $0_{g}^{-}$ state of $^{85}$Rb$_{2}$, bound-bound spectroscopy has determined the energies of the $v=0$ to 14 vibrational states \cite{MHH09}, and photoassociative spectroscopy has determined the binding energies of the vibrational levels between $v=184$ and 171 \cite{BTH99}.
Combining the two data sets leaves roughly 150 unknown vibrational energies.
Many more vibrational energies, including those studied here, can be identified in the photoassociation spectra of ref \cite{MCH93}, but in this paper binding energies were not given. 

In 2008, C\^{o}t\'{e} and colleagues suggested that Feshbach resonances could be used to systematically enhance PA and optimize the production of cold molecules \cite{PGC08}.
This technique, Feshbach optimized photoassociation (FOPA) modifies the wavefunction of the colliding atoms to optimize the creation of molecules in the desired vibrational states. 
Additional calculations showed that FOPA may be used to create heternuclear LiCs molecules \cite{GC14}, to study the time variation of the electron-to-proton mass ratio \cite{GC14}, and to test the unitarity limit \cite{PC09}.
In fact, the first observation of a Feshbach resonance in ultracold collisions was made by observing the enhancement of the photoassociation rate \cite{CFH98}.

In this work, we observe Feshbach optimized photoassociation of $^{85}$Rb$_{2}$ formed in deeply-bound vibrational levels of the $0_{g}^{-}$ electronic potential that connects to the 5$^{2}$S$_{1/2}$+5$^{2}$P$_{1/2}$ separated atoms limit. 
We focus on atoms that collide in the entrance channel $|F=2, m_{F}=-2 \rangle$ + $|F=2, m_{F}=-2 \rangle$, which couples at small internuclear separation to other hyperfine states with total angular momentum projection quantum number $M_{F}=-4$.
These atoms form a broad Feshbach resonance centered at an external magnetic field strength of 155 G.
Specifically, we measure the binding energies of the $v=127$, 140, and 150 vibrational states. 
We also measure the magnetic field position and width of each FOPA resonance, and compare the results to those from to a multi-channel $s$-wave scattering calculation.

Previous experimental and theoretical studies have investigated Feshbach resonance and photoassociation.
The first observation of a Feshbach resonance using PA also demonstrated FOPA for the $v=181$ state of the $0_{g}^{-}$ potential in $^{85}$Rb$_{2}$ \cite{CFH98}.
A later study on $^{133}$Cs$_{2}$ \cite{THJ03} used a Feshbach resonance to enhance the photoassociation rate into many deeply-bound vibrational levels of the $0_{g}^{-}$ excited molecular potential.
Feshbach optimized photoassociation studied the unitarity limit in Li$_{2}$ at low laser intensity \cite{JDW08}. 
Most recently, Semczuk $et$ $al.$ \cite{SLG13} used FOPA as part of the process to obtain a series of previously unresolved vibrational levels ($v=20 - 26$) in the $1^{3} \Sigma _{g}^{+}$ excited molecular potential of $^{6}$Li$_{2}$. 

\section{Theory}

\begin{figure}[H]
\begin{center}
\includegraphics[width=7.7 cm]{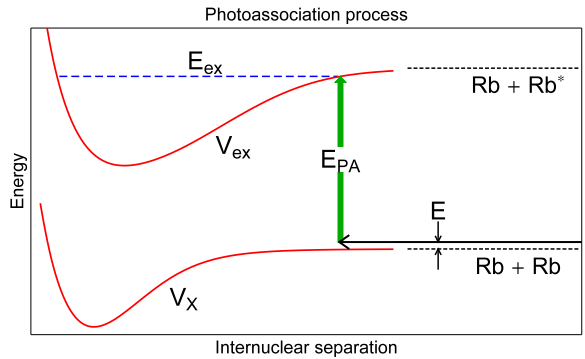}
\caption{\label{pa} (Color online) A diagram of the photoassociation process.  Two rubidium atoms Rb + Rb in a trapped, ultracold gas collide with relative kinetic energy $E$ above a ground molecular potential $V_X$. 
During the collision, a photon of energy $E_{PA}$ promotes the pair into a bound vibrational state of energy $E_{ex}$ in the excited molecular potential $V_{ex}$ that dissociates to a ground-state atom and singly excited atom (Rb+Rb*).
 }
\end{center}
\end{figure}

Photoassociaton begins with two free colliding atoms under the influence of a ground-state molecular potential. 
These atoms are associated into an excited-state molecule by a photon \cite{JTL06} (Figure~\ref{pa}).  
The excited molecule spontaneously decays into a molecule in the ground state or into two free atoms.  
This provides a method to detect PA.  
The atoms are initially confined in a trap, but the byproducts usually are not.  
The molecule rarely experiences the same trapping forces as the atoms, and the free atoms often have sufficient energy to escape the trap.
The rate of photoassociation can be estimated by the Franck-Condon factor, the space-integral of the product of the ground- and excited-state wavefunctions. 
Because of the asymmetry of the excited-state confining potential, the amplitude of the wavefunction is concentrated near the outer turning point. 
Since the free wavefunctions have large amplitudes at large atomic separation, the Frank-Condon factors are largest for the highest-lying vibrational levels of the excited molecular potential.  
The photoassociation rate generally decreases for more deeply-bound vibrational levels, which have turning points at shorter internuclear separation.  
Also, oscillations in this rate can reflect the oscillations of the ground-state wavefunction amplitude \cite{AMG96}

\begin{figure}
\begin{center}
\includegraphics[width=7.7 cm]{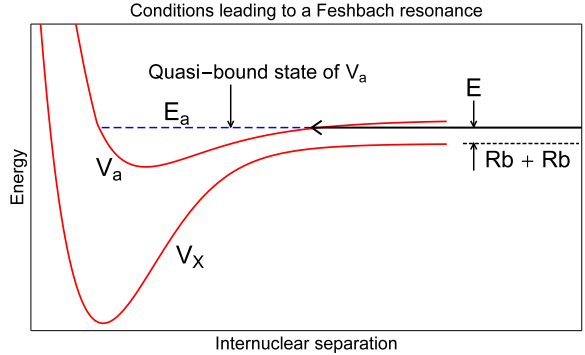}
\caption{\label{fr} (Color online) Two atoms initially in a particular separated atoms state collide with relative kinetic energy $E$. A magnetic field shifts the open channel potential V$_{X}$, such that the collision energy is equal to a bound-state energy, $E_a$ of the closed channel V$_{a}$. This leads to coupling of the two channels and enhancement of the initial scattering wave function amplitude at small internuclear separation.  }
\end{center}
\end{figure}

The electronic molecular potential of two ground-state rubidium atoms has multiple dissociation energies corresponding to different combinations of hyperfine energies of the constituent atoms.  
Their relative dissociation energies can be adjusted relative to each other by an external magnetic field.  
Thus, if two atoms are colliding along one interaction potential, their collision energy can be made equal to the energy of a molecular state with a slightly higher dissociation energy by applying the appropriate magnetic field (Figure~\ref{fr}).  
This Feshbach resonance \cite{KGJ06} couples the molecular potentials and dramatically alters the collision wavefunction, especially at small internuclear separation.  
This, in turn, changes the photoassociation rate into the electronically excited molecular potential.  
Utilizing this process, molecular states can be formed that would otherwise be inaccessible.
\section{Experiment}

\begin{figure}
\begin{center}
\includegraphics[width=7.7 cm]{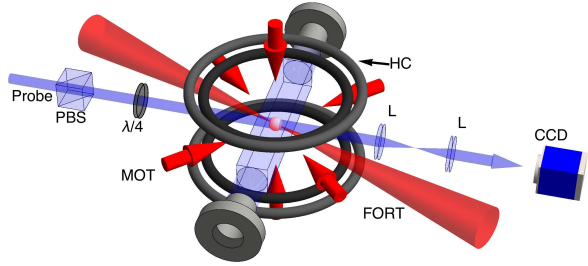}
\caption{\label{fort} (Color online) The experimental set-up for the far-off resonance trap (FORT). Six orthogonal beams (red arrows) overlap at the zero of a magnetic quadrupole field to form a magneto-optical trap (MOT). A focused Ti:S laser (red cones) creates the FORT. Helmholtz coils (HC) produce the magnetic field necessary for the Feshbach resonance. The probe laser passes through the trapped atoms and the atom density is measured by the absorption of the probe laser measured by the CCD camera. }
\end{center}
\end{figure}

Figure~\ref{fort} shows our experimental set-up and is similar to that in \cite{AKM15}.
We begin with atoms in a magneto-optical trap (MOT) \cite{RPC87} that contains $\sim10^{8}$ atoms with a density of $\sim$$10^{10}$ cm$^{-3}$, a 1/$e^{2}$ radius of 2 mm, and a temperature of 50 $\mu$K.
We use two external cavity diode lasers (ECDL) for the trapping and repump beams.
Each laser is frequency stabilized using a dichroic atomic vapor laser lock \cite{MHT07}.  
The trapping laser is detuned $\simeq$15 MHz to the red of the $5^{2}$S$_{1/2}(F=3) \rightarrow 5^{2}$P$_{3/2}(F^{\prime}=4)$ atomic transition frequency.
The laser is amplified by a tapered amplifier in the master oscillator power amplifier configuration.
The output is re-shaped and spatially filtered by an angle-polished polarization-maintaining single mode optical fiber. 
At the output of the fiber, the beam has a power of 175 mW and is telescoped to a 1/$e^{2}$ diameter of 2.5 cm.
The repump laser is locked on resonance to the $5^{2}$S$_{1/2}(F=2) \rightarrow 5^{2}$P$_{3/2}(F^{\prime}=3)$ atomic transition frequency.
The repump beam has a power of 10 mW and a 1/$e^{2}$ diameter of 2.5 cm.
The intensity of the repump laser is controlled by an acousto-optical modulator.
The repump laser overlaps the trapping laser, and the resulting beam is split 6 ways.
Three orthogonal circularly polarized beams intersect at the center of the vacuum cell, and are co-propagating with the other three beams with opposite polarization. 
A quadrupole magnetic field gradient created by a pair of anti-Helmholtz coils that is 9 G/cm along the strong axis, and that can be turned off in 10 $\mu$s. 
The probe laser is an ECDL locked to the $5^{2}$S$_{1/2}(F=3) \rightarrow 5^{2}$P$_{3/2}(F^{\prime}=4)$ transition with $\sigma^{+}$ polarization provides absorption images on a CCD camera. 
A Titanium Sapphire (Ti:S) laser (Spectra Physics Millennium 3900s) with 2.1 W of power and a $\sim$40 GHz linewidth creates the confining potential for a far-off resonant optical dipole trap (FORT) \cite{MCH93b}.
The laser is focused down to a 1/$e^{2}$ radius of 30 $\mu$m at the position of the MOT.
The samples in the FORT routinely contain $5 \times 10^{5}$ atoms with a density of $\sim$10$^{10}$ cm$^{-3}$ and a temperature between 200 and 300 $\mu$K.

Wire coils in a Helmholtz configuration create the uniform field for the Feshbach resonance. 
The coils are made from 24 wraps of copper tubing around a Phenolic base with an average coil radius of 11.2 cm.
Roughly 65 A is needed to produce the magnetic field for the Feshbach resonance.
The coils are powered by a Walker Scientific HS 52400-4SS power supply.
The supply can deliver up to 400 Amps with a stability in 1 part to the $10^{6}$.
The supply is controlled through an external  DC voltage, and can turn on/off in roughly 30 ms.

\begin{figure}
\begin{center}
\includegraphics[width=7.7 cm]{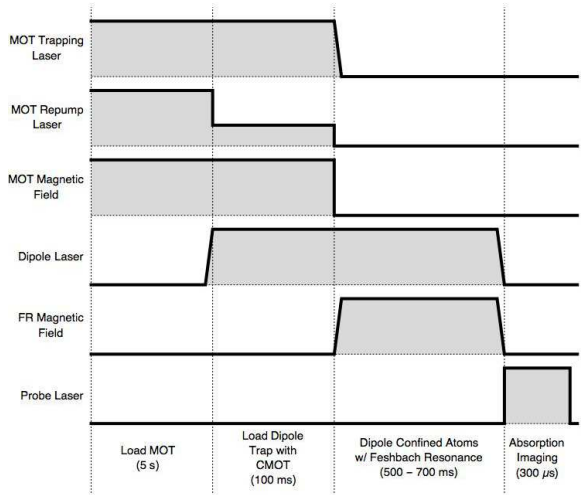}
\caption{\label{FOPA_timing} The timing sequence to induce Feshbach optimized photoassociation. The magneto-optical trap is loaded, and the atoms are transferred to a far off-resonance trap tuned to a photoassociation resonance. The photoassociation rate is enhanced by a Feshbach resonance, the far-off resonance trap is turned off, and the shadow of the atoms is imaged on a CCD camera. The vertical axis indicates the quantity being off or on, and in the case of the MOT repump laser, being at a lower intensity. The vertical axis is not to scale. }
\end{center}
\end{figure}

Figure~\ref{FOPA_timing} shows the timing sequence for the experiment.
The sequence begins by loading the MOT for 5 s. 
Next, the Ti:S laser is directed through the atom cloud, the MOT trapping laser is detuned 120 MHz to the red, and the repump intensity is reduced to 0.5 $\mu$W/cm$^{2}$.
This stage optically pumps the atoms into the $F=2$ ground state to prevent atom loss from hyperfine changing collisions and lasts for 100 ms.
The quadrupole magnetic field and repump laser are turned off, and the trapping laser is shuttered 1.5 ms later.
We turn on the Feshbach coils to create a constant magnetic field for the Feshbach resonance in 30 ms. 
The atoms remain in the FORT and under the influence of the magnetic field for 560 to 660 ms.
The same laser induces the photoassociation resonance. 
The magnetic field is turned off and the Ti:S laser is shuttered.
The repump laser flashes on for 100 $\mu$s to drive the atoms into the $F=3$ ground state.
The probe laser is turned on for 350 $\mu$s to image the atoms.
Five hundred ms later, the probe laser flashes on for 350 $\mu$s to image the probe beam without the presence of ultracold atoms.
Next, the camera takes another picture for 350 $\mu$s to image the background light.
This process is repeated for different magnetic field values separated by 1 G increments.
The order of the magnetic field values is randomized to minimize temporal effects such as laser drift appearing as signal.
\section{Results/Discussion}

\begin{figure}
\begin{center}
\includegraphics[width=7.7 cm]{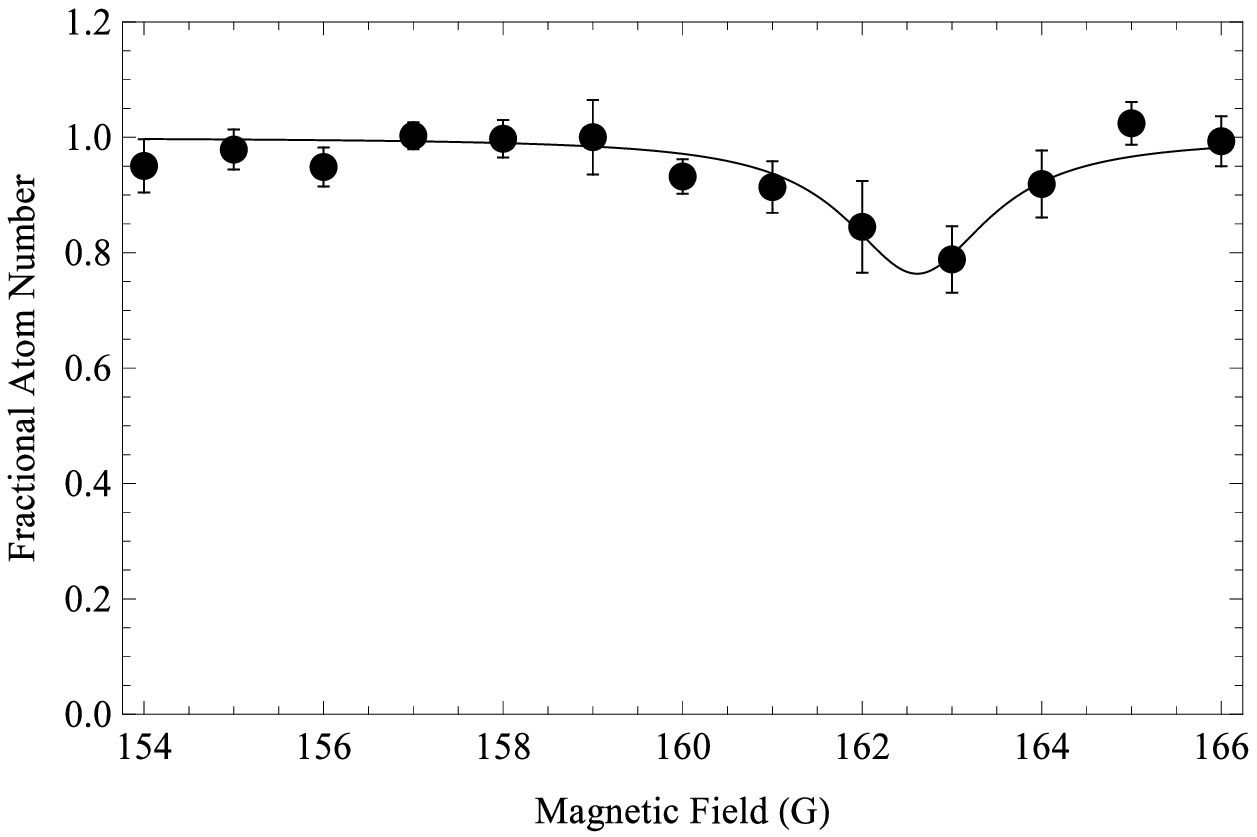}
\caption{\label{12323_exp}  The circles represent the fractional number of atoms remaining in the trap for each value of the magnetic field at a trapping laser frequency of 12323 cm$^{-1}$, which excites the colliding atoms into the $\nu=127$ state of the $0^{-}_{g}$ excited molecular potential. The curve is a Lorentzian fit to the data.   }
\end{center}
\end{figure}

\begin{figure}
\begin{center}
\includegraphics[width=7.7 cm]{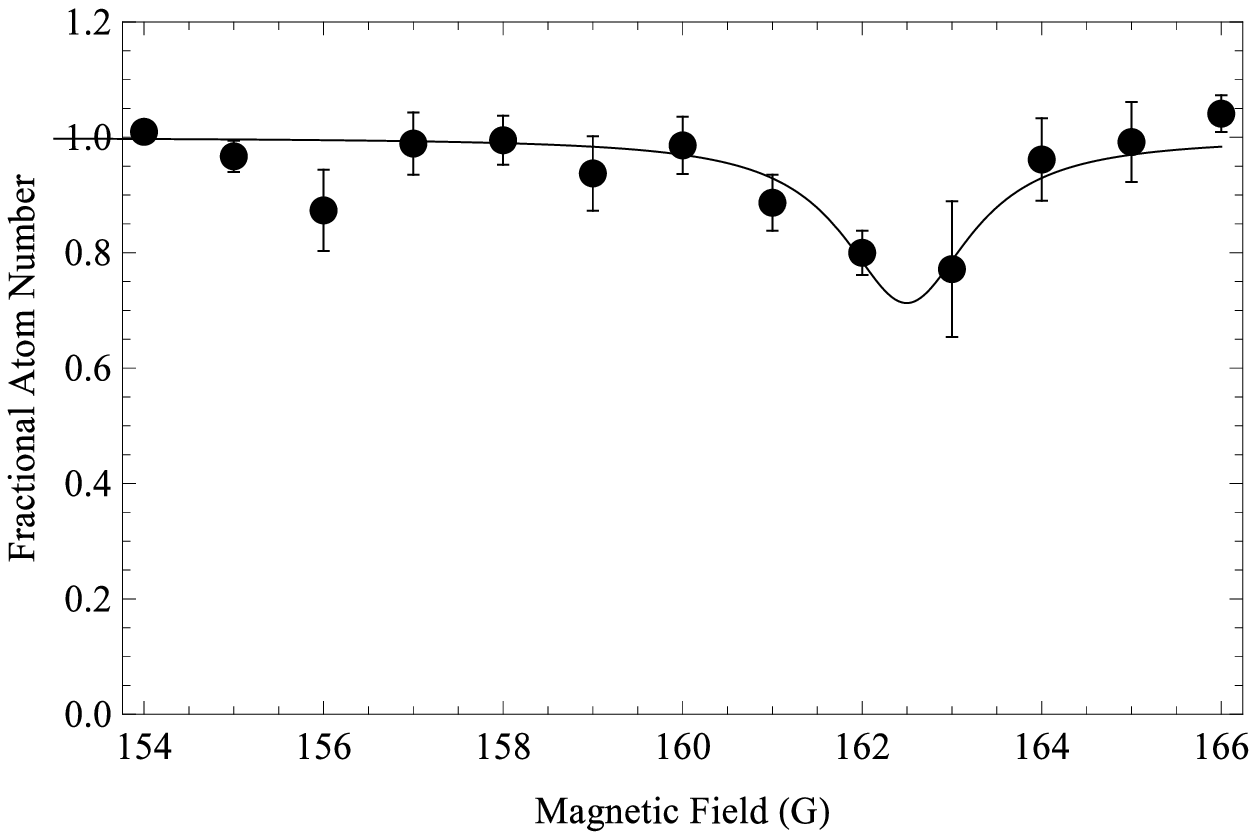}
\caption{\label{12425_exp}  The same as Fig.~\ref{12323_exp} with a trapping laser frequency of 12425 cm$^{-1}$, which excites the colliding atoms into the $\nu=140$ state of the $0^{-}_{g}$ excited molecular potential.}
\end{center}
\end{figure}

\begin{figure}
\begin{center}
\includegraphics[width=7.7 cm]{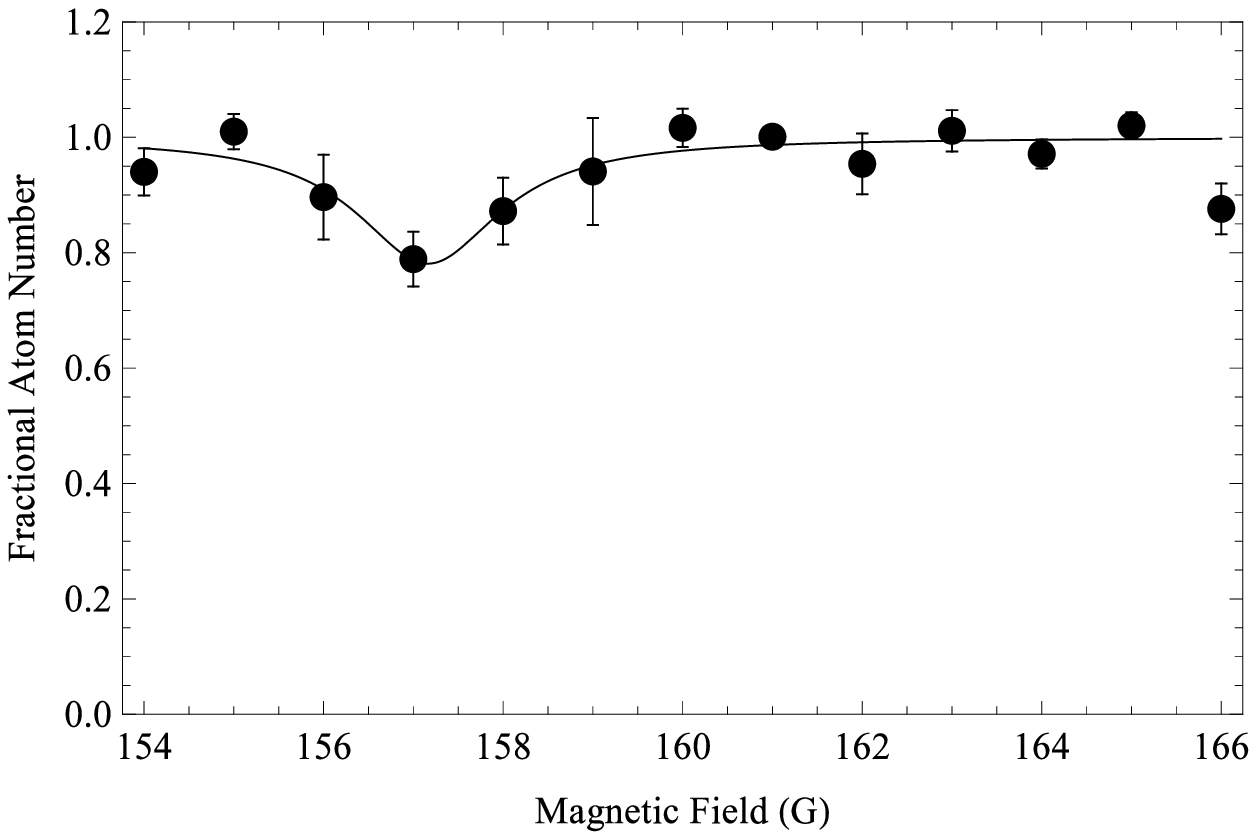}
\caption{\label{12483_exp}   The same as Fig.~\ref{12323_exp} with a trapping laser frequency of 12483 cm$^{-1}$, which excites the colliding atoms into the $\nu=150$ state of the $0^{-}_{g}$ excited molecular potential.  }
\end{center}
\end{figure}

We observe three separate FOPA resonances in $^{85}$Rb$_{2}$.
These are shown in Figures ~\ref{12323_exp},~\ref{12425_exp}, and~\ref{12483_exp}.
Based on theoretical calculations of binding energies, these levels correspond to the vibrational levels $v=127$, 140, and 150 of the $0_{g}^{-} (5$S$_{1/2}+5$P$_{1/2}$) state, with measured binding energies of 256 cm$^{-1}$, 154 cm$^{-1}$, and 96 cm$^{-1}$ below the 5$^{2}$S$_{1/2}$+5$^{2}$P$_{1/2}$ separated-atoms limit.
The dissociation energy is 3186.562 cm$^{-1}$.
The frequency width of the photoassociation laser is 1.2 cm$^{-1}$, which determines the uncertainty in the measured vibrational energy.
We cannot resolve molecular rotational levels, and at the measured temperature we may be exciting from an initial state composed of several partial waves into quantum states with varying rotational quantum numbers. 
Absent the Feshbach resonance we do not observe photoassociative loss in our system.
The experiment is repeated with the FORT laser at a frequency where we do not expect a photoassociation resonance. 
We do not see any change in PA as a function of applied field strength $B$, indicating that the results in  Figures ~\ref{12323_exp}, ~\ref{12425_exp}, and ~\ref{12483_exp} are due to FOPA and not to an increase of two-body or three-body loss due to collisions.
This is consistent with expected loss rates at our densities \cite{RCC00}.
The first observed resonance is at a trapping laser frequency of 12323 cm$^{-1}$, corresponding to a transition into the $v=127$ vibrational level with a binding energy of $256 \pm 0.6$ cm$^{-1}$.
Figure~\ref{12323_exp} shows the fractional number of atoms remaining in the FORT as a function of the magnetic field, $B$, as it is swept through the Feshbach resonance.
The number of atoms remaining in the FORT is calculated by summing the values of the pixels in the absorption images of the trapped atoms.
The resulting data is normalized to the background number of atoms when the magnetic field is off resonance.               
Figure 5 contains an average of 9 data sets. 
The error bars represent statistical uncertainties. 

The data is fit to a Lorentzian function given by,
\begin{equation}
	1- \frac{a(\Delta/2)^2}{(B-B_{0})^2+(\Delta/2)^{2}},
\end{equation}
where the fitting parameters are $a$, the amplitude of the peak; $\Delta$, the full width at half the maximum (FWHM); and $B_{0}$, the central magnetic field position of the peak \cite{CFH98}.
A reduced chi-square analysis gives a FWHM of $\Delta=3^{+4}_{-2}$ G and a central magnetic field value of $B_{0} = 163^{+2}_{-1}$ G where the errors are statistical.
There is an additional systematic error of $\pm$3 G in $B_{0}$ from the uncertainty in the calibration of the Helmholtz coils.
The reduced chi-squared of the best fit is 0.9. 

The second observed resonance is at a trapping laser frequency of $12425 \pm 0.6$ cm$^{-1}$, corresponding to a transition into the $v=140$ vibrational level with a binding energy of $154 \pm 0.6$ cm$^{-1}$.
Figure~\ref{12425_exp} shows the fractional number of atoms remaining in the FORT as a function of the magnetic field, $B$.
Figure 6 contains an average of 5 data sets and the error bars represent the statistical uncertainty in the data. 
A similar reduced chi-square analysis results in a FWHM of $\Delta=2^{+3}_{-1}$ G and a central magnetic field value of $B_{0} = 162 \pm 1 \pm 3$ G (where the first error is statistical and the second systematic).
The reduced chi-squared of the best fit is 1.1. 

The final observed resonance is at a trapping laser frequency of 12483 cm$^{-1}$, corresponding to a transition into the $v=150$ vibrational level with a binding energy of $96 \pm 0.6$ cm$^{-1}$.
Figure~\ref{12483_exp} shows the fractional number of atoms remaining in the FORT as a function of the magnetic field, $B$, as it is swept through the Feshbach resonance.
Figure 6 contains an average of 5 data sets. 
A reduced chi-square analysis gives a FWHM of $\Delta=2^{+2}_{-1}$ G and a central magnetic field value of $B_{0} = 157 \pm 1 \pm 3$ G.
The reduced chi-squared of the best fit is 1.8. 

We have performed close-coupling calculations of the Feshbach resonance and from the results, have calculated the associated photoassociation rates for doubly spin-polarized Rb atoms excited into the $0_{g}^{-}$ molecular potential.
The calculations take into account only $s$-wave collisions and transitions into the $J=0$ rovibrational state. 
For $v=127$, our calculation gives a resonance position that varies between 158-163 G and a resonance width that varies between 9-5 G for temperatures between 10 $\mu$K and 200 $\mu$K. 
Our theoretical model shows that as the temperature increases from 10 $\mu$K, the Feshbach resonance peak moves to higher magnetic fields and the resonance width narrows. 
Previously, a similar measurement of a $0_{g}^{-}$ bound molecular vibrational state with 5.9 cm$^{-1}$ binding energy found a resonance peak of $B_{0}=166.6 \pm 7$ G and a width of $\Delta=5.9 \pm 2.1$ G \cite{CFH98}.

Our results agree with those from the previous experiment \cite{CFH98} to within one standard deviation, as well as those from our theoretical calculation \cite{D15} with the exception of the width of $v=150$ Feshbach resonance, which agrees at the 2$\sigma$ level to the theoretical width. 
The measured values are consistently lower than those from previous experiment \cite{CFH98} and theory \cite{D15}.
This discrepancy may be due to the lower signal-to-noise in our experiment compared to \cite{CFH98}, and that our initial system is not doubly-spin polarized (all the atoms in the $|F=2, m_{F}=-2 \rangle$ state).
Also, excitations into rotational states with narrower widths may contribute to the signal.

\section{Conclusions}

We have observed three Feshbach optimized photoassociation resonances in $^{85}$Rb$_{2}$ and measure binding energies of  $256 \pm 0.6$ cm$^{-1}$, $154 \pm 0.6$ cm$^{-1}$, and $96 \pm 0.6$ cm$^{-1}$ below the 5$^{2}$S$_{1/2}$+5$^{2}$P$_{1/2}$ separated atoms limit.
We measure the Feshbach resonance enhancement to occur at $B_{0} = 163^{+2}_{-1}  \pm 3$ G ($v=127$),  $B_{0} = 162 \pm 1 \pm 3$ G ($v=140$), and $B_{0} = 157 \pm 1 \pm 3$ G ($v=150$), and corresponding widths of $\Delta=3^{+4}_{-2}$ G, $\Delta=2^{+3}_{-1}$ G, and $\Delta=2^{+2}_{-1}$ G.
Our results are consistent with theory and previous experimental work.
We do not observe PA without the Feshbach resonance present, indicating the usefulness of FOPA in determining vibrational energies of excited molecular potentials. 

\section{Acknowledgments}

We thank Gregory Parker for insightful discussions. 
This work is supported by the Office of Naval Research Contract No. N00014-02-1-0601.

\end{document}